\documentclass[prl,twocolumn,preprintnumbers,amsmath,aps]{revtex4}
\usepackage{graphicx,hyperref}% Include figure files
\usepackage{dcolumn}% Align table columns on decimal point
\usepackage{bm}% bold math
\usepackage{subfigure}% bold math

\newcommand\gk{\Gamma_k}
\newcommand\rk{R_k(q)}

\newcommand\Min{{\hbox{\begin{tiny}m\end{tiny}}}}

\newcommand\ie{{\it i.e. }}

\newcommand\vect[1]{ \boldsymbol{ #1}}
\def\nbC{{\mathchoice {\setbox0=\hbox{$\displaystyle\rm C$}%
\hbox{\hbox to0pt{\kern0.4\wd0\vrule height0.9\ht0\hss}\box0}}
{\setbox0=\hbox{$\textstyle\rm C$}\hbox{\hbox
to0pt{\kern0.4\wd0\vrule height0.9\ht0\hss}\box0}}
{\setbox0=\hbox{$\scriptstyle\rm C$}\hbox{\hbox
to0pt{\kern0.4\wd0\vrule height0.9\ht0\hss}\box0}}
{\setbox0=\hbox{$\scriptscriptstyle\rm C$}\hbox{\hbox
to0pt{\kern0.4\wd0\vrule height0.9\ht0\hss}\box0}}}}
\def\nbQ{{\mathchoice {\setbox0=\hbox{$\displaystyle\rm
Q$}\hbox{\raise
0.15\ht0\hbox to0pt{\kern0.4\wd0\vrule height0.8\ht0\hss}\box0}}
{\setbox0=\hbox{$\textstyle\rm Q$}\hbox{\raise
0.15\ht0\hbox to0pt{\kern0.4\wd0\vrule height0.8\ht0\hss}\box0}}
{\setbox0=\hbox{$\scriptstyle\rm Q$}\hbox{\raise
0.15\ht0\hbox to0pt{\kern0.4\wd0\vrule height0.7\ht0\hss}\box0}}
{\setbox0=\hbox{$\scriptscriptstyle\rm Q$}\hbox{\raise
0.15\ht0\hbox to0pt{\kern0.4\wd0\vrule height0.7\ht0\hss}\box0}}}}
\def\nbT{{\mathchoice {\setbox0=\hbox{$\displaystyle\rm
T$}\hbox{\hbox to0pt{\kern0.3\wd0\vrule height0.9\ht0\hss}\box0}}
{\setbox0=\hbox{$\textstyle\rm T$}\hbox{\hbox
to0pt{\kern0.3\wd0\vrule height0.9\ht0\hss}\box0}}
{\setbox0=\hbox{$\scriptstyle\rm T$}\hbox{\hbox
to0pt{\kern0.3\wd0\vrule height0.9\ht0\hss}\box0}}
{\setbox0=\hbox{$\scriptscriptstyle\rm T$}\hbox{\hbox
to0pt{\kern0.3\wd0\vrule height0.9\ht0\hss}\box0}}}}
\def\nbS{{\mathchoice
{\setbox0=\hbox{$\displaystyle     \rm S$}\hbox{\raise0.5\ht0%
\hbox to0pt{\kern0.35\wd0\vrule height0.45\ht0\hss}\hbox
to0pt{\kern0.55\wd0\vrule height0.5\ht0\hss}\box0}}
{\setbox0=\hbox{$\textstyle        \rm S$}\hbox{\raise0.5\ht0%
\hbox to0pt{\kern0.35\wd0\vrule height0.45\ht0\hss}\hbox
to0pt{\kern0.55\wd0\vrule height0.5\ht0\hss}\box0}}
{\setbox0=\hbox{$\scriptstyle      \rm S$}\hbox{\raise0.5\ht0%
\hboxto0pt{\kern0.35\wd0\vrule height0.45\ht0\hss}\raise0.05\ht0%
\hbox to0pt{\kern0.5\wd0\vrule height0.45\ht0\hss}\box0}}
{\setbox0=\hbox{$\scriptscriptstyle\rm S$}\hbox{\raise0.5\ht0%
\hboxto0pt{\kern0.4\wd0\vrule height0.45\ht0\hss}\raise0.05\ht0%
\hbox to0pt{\kern0.55\wd0\vrule height0.45\ht0\hss}\box0}}}}
\def\nbZ{{\mathchoice {\hbox{$\sf\textstyle Z\kern-0.4em Z$}}
{\hbox{$\sf\textstyle Z\kern-0.4em Z$}}
{\hbox{$\sf\scriptstyle Z\kern-0.3em Z$}}
{\hbox{$\sf\scriptscriptstyle Z\kern-0.2em Z$}}}}
\begin{document}

\title{A unified picture of ferromagnetism, quasi-long range order and criticality in random
  field models}

\author{Matthieu Tissier}
%\email{tissier@lptl.jussieu.fr}
\affiliation{LPTMC, Universit\'e Pierre et Marie Curie, bo\^ite 121, 4
  Pl. Jussieu, 75252 Paris c\'edex 05, France} 

\author{Gilles Tarjus}
%\email{tarjus@lptl.jussieu.fr}
\affiliation{LPTMC, Universit\'e Pierre et Marie Curie, bo\^ite 121, 4
  Pl. Jussieu, 75252 Paris c\'edex 05, France}

%\date{\today}

\begin{abstract}
  By applying the recently developed nonperturbative functional
  renormalization group (FRG) approach, we study the interplay between
  ferromagnetism, quasi-long range order (QLRO) and criticality in the
  $d$-dimensional random field $O(N)$ model in the whole ($N$, $d$)
  diagram. Even though the "dimensional reduction" property breaks
  down below some critical line, the topology of the phase diagram is
  found similar to that of the pure $O(N)$ model, with however no
  equivalent of the Kosterlitz-Thouless transition. In addition, we
  obtain that QLRO, namely a topologically ordered "Bragg glass"
  phase, is absent in the $3$--dimensional random field XY model. The
  nonperturbative results are supplemented by a perturbative FRG
  analysis to two loops around $d=4$.
\end{abstract}

%\pacs{11.10.Hi, 75.40.Cx}

\maketitle

How the phase behavior and ordering transitions of a system are
affected by the presence of a weak random field remains in part an
unsettled problem. Heuristic and rigorous arguments show that the
lower critical dimension below which no long-range order is possible
is $2$ for the random field Ising model (RFIM)\cite{imry75,
  bricmont87} and $4$ for models with a continuous symmetry (RF$O(N)$M
with $N>1$)~\cite{imry75,larkin70,aizenman89}. However, this leaves
aside two questions: first, the nature of the critical behavior in
random field models, a question connected to the breakdown of the
so-called "dimensional reduction" (DR) property that relates the
critical exponents of the RF$O(N)$M to those of the pure $O(N)$ model
in two dimensions less~\cite{nattermann98}; and second, the possible
occurence of a low-temperature phase with quasi-long range order
(QLRO), \ie, a phase characterized by no magnetization and a power-law
decrease of the correlation fuctions, in models with a continuous
symmetry~\cite{giamarchi98,feldman01}. Progress has been made to better
circumscribe this latter point. It has indeed been shown that QLRO is
absent for $N=2$ when disorder is strong and for $N>3$ for arbitrarily
weak random field~\cite{feldman00}; but this still keeps
open the cases $N=2,3$ in the physical dimensions $d=2,3$.

Those questions are important because, on top of purely theoretical
motivations, they concern the behavior of the known experimental
realizations of random field models. This is the case for instance of
vortex lattices in disordered type-II
superconductors~\cite{giamarchi98,blatter94}. In such systems, the
randomly pinned lattice of vortices can be mapped onto an "elastic
glass model"~\cite{giamarchi98,blatter94}, whose simplest realization
is the $N=2$ RF$XY$M. The occurence of a phase with QLRO, termed
"Bragg glass", has been predicted for the $3-d$ version of the
model~\cite{giamarchi98}. Further theoretical support for this
prediction has been given by a Monte Carlo simulation of the
RF$XY$M~\cite{gingras96} and by analyses of the energetics of
dislocation loops~\cite{giamarchi98,fisher97}.

In this letter, we apply our recently developed nonperturbative FRG
approach of the RF$O(N)$M~\cite{tarjus04} to provide a unified picture of
ferromagnetism, QLRO and criticality in the whole ($N$, $d$) diagram.
We find that below a critical value $N_{c}=2.83..$ and for $d<4$ the
model has a transition to a QLRO phase, both this phase and the
transition being governed by zero-temperature nonanalytic fixed
points (FPs). The transition disappears below a lower critical dimension
which we find around $3.8$ for $N=2$. Therefore, contrary to what is
usually believed, no QLRO and no Bragg glass phase exist in the $3-d$
RF$XY$M. We supplement our nonperturbative, but of course approximate,
results by a perturbative FRG analysis to two loops in $d=4+\epsilon$.
The present approach allows us to discuss the DR
property and its breakdown. We find in particular that the topology of
the ($N$, $d$) phase diagram of the RF$O(N)$M is similar to that of
the pure $O(N)$ model, with however no equivalent of the ($N=2$,
$d=2$) Kosterlitz-Thouless (KT) transition. 

Our starting point is the standard effective hamiltonian for the
RF$O(N)$M in $d$ dimensions with an $N$-component field $\vect\chi(x)$
and uncorrelated random fields taken from a Gaussian distribution with
zero mean and variance $\Delta$. After introducing $n$ replicas in
order to perform the average over quenched disorder (taking at the end
the limit $n\to0$), it can be rewritten as:
\begin{equation}
  \label{eq_action_replicated}
  \begin{split}
    S[\{\vect\chi_a\}]=\int_x \Bigg\{&\frac1{2T}
    \sum_{a=1}^n \bigg[ \partial\vect{\chi}_a \cdot
    \partial\vect{\chi}_a + \tau \vect{\chi}_a^2 +\frac
    u{12}\vect{\chi}_a^4\bigg]\\&- \frac\Delta{2T^2}
    \sum_{a,b=1}^n\vect{\chi}_a.\vect{\chi}_b -\sum_{a=1}^n
    \vect{J}_a.\vect{\chi}_a\Bigg\},
  \end{split}
\end{equation}
where we have introduced sources acting on each replica separately,
which therefore explicitly break the permutation symmetry between the
replicas.

We apply in this work the nonperturbative FRG formalism for the
RF$O(N)$M recently proposed by us~\cite{tarjus04}. To keep the presentation
short but sufficiently self-contained, we first sketch the main steps
of the approach. It is based on an exact RG equation for the effective
average action $\gk[\{\vect\phi_a\}]$. $\gk$ interpolates between the
bare action, eq.(\ref{eq_action_replicated}), and the usual effective
action (i.e., the Legendre transform of the partition function
associated with eq.(\ref{eq_action_replicated})) as the running scale
$k$ moves from microscopic ($k=\Lambda$) to macroscopic ($k\to0$)
scale. It is built by integrating out fluctuations with momenta
larger than $k$ and its flow obeys an exact equation,
\begin{equation}
  \label{eq_flot_exact}
  \partial_t\gk[\{\vect\phi_a\}]=\frac 12\text{Tr}\{ \partial_t\rk(
  \gk^{(2)}[\{\vect\phi_a\},q] + \openone \rk)^{-1}\}
\end{equation}
where $\partial_t$ is a derivative with respect to $t=\ln(k/\Lambda)$,
$\gk^{(2)}$ is the tensor formed by the second functional derivatives
of $\gk$ with respect to the fields $\vect\phi_a(q)$, $\openone$ is
the unit tensor, and the trace involves an integration over momenta as
well as a sum over replica indices and $N$-vector components. $\rk$ is
the infrared cutoff introduced to suppress the low-momentum modes.  In
practice, solving eq.(\ref{eq_flot_exact}) numerically requires the
introduction of approximation schemes which amount to truncate the
functional form of $\gk$. Guided by the physics of the problem at hand
and by the mounting work on the method~\cite{berges02},
one can then formulate a nonperturbative RG description. For the
RF$O(N)$M we have argued that the two main ingredients allowing to
study the long-distance physics are (i) the derivative
expansion, which approximates the momentum dependence of the (1PI)
vertex functions~\cite{berges02}, and (ii) the expansion in increasing
number of free replica sums, which is equivalent to include
increasing-order cumulants of the renormalized distribution of the
quenched disorder~\cite{tarjus04}.

The simplest nonperturbative FRG
description of the RF$O(N)$M relies on the following truncation:
\begin{equation}
  \label{eq_trunc_1}
  \begin{split}
    \gk[\{\vect \phi_a\}]=\int_x\Bigg\{&\sum_{a=1}^n\left( \frac 12
    Z_{\Min,k} \partial\vect \phi_a\cdot\partial\vect \phi_a+U_k(\vect\phi_a)
    \right)\\
    -&\frac 12 \sum_{a,b=1}^nV_k(\vect\phi_a,\vect\phi_b)\Bigg\}
  \end{split}
\end{equation}
with one single wavefunction renormalization for all fields,
$Z_{\Min,k}$, evaluated at the field configurations that minimize the
1-replica potential $U_k$ (pseudo first-order derivative
expansion~\cite{berges02}); the 1-replica $U_k$ and 2-replica $V_k$
parts of the effective potential are obtained from $\gk$ taken for
uniform fields. Inserting eq.(\ref{eq_trunc_1}) into
eq.(\ref{eq_flot_exact}), leads to coupled partial differential
equations for the functions $U_k(\tilde{\rho})$ and
$V_k(\tilde{\rho},\tilde\rho',z)$, where $\tilde{\rho}=\frac 12
\vect\phi^2$ and
$z=\vect\phi\cdot\vect\phi'/\sqrt(4\tilde\rho\tilde\rho')$, and to a
running anomalous dimension defined as $\eta_k=-\partial_t\log
Z_{\Min,k}$. The sought FPs being at zero temperature, one also
introduces a running temperature $T_k$ and the associated exponent
$\theta_k=\partial_t\log T_k$. This is most conveniently done by
defining a renormalized disorder strength
$\Delta_{\Min,k}=(2\tilde\rho_{\Min,k})^{-1}\partial_z
V_k(\tilde\rho_{\Min,k}, \tilde\rho_{\Min,k}, z=1)$, where
$\tilde\rho_{\Min,k}$ corresponds to the minimum of
$U_k(\tilde{\rho})$. Defining then $T_k$ as $Z_{\Min,k}
k^2\Delta/(\Lambda\Delta_{\Min,k})$, one can see that it reduces to
the bare temperature $T$ at the microscopic scale $\Lambda$.

The flow equations can be expressed in a scaled form by introducing
renormalized dimensionless quantities,
$u_k(\rho)=T_kk^{-d}U_k(\tilde{\rho})$, $v_k(\rho,\rho',z)=
T_k^2k^{-d}V_k(\tilde{\rho},\tilde\rho',z)$,
$\rho=Z_{\Min,k}T_kk^{-(d-2)}\tilde{\rho}$: see eqs. (5,6) of
ref.~\cite{tarjus04}. An expression for $\eta_k$ is derived by considering
the flow of the transverse component of $\gk^{(2)}$ evaluated for
uniform fields and zero momentum:
\begin{equation}
  \label{eq_etab}
  \begin{split}
    \eta_k=&8C_d/d\big\{u_{\rho \rho} [v_z-4 \rho_{\Min} v_{\rho
      \rho'}]m_{3,1}^d(2\rho_{\Min}u_{\rho
      \rho} ,0)\\
    &+2 u_{\rho \rho}^2 v_z[m_{2,3}^d(2\rho_{\Min} u_{\rho
      \rho},0)+m_{3,2}^d(2 \rho_{\Min}u_{\rho \rho} ,0)] \big\}
  \end{split}
\end{equation}
where $C_d^{-1}=2^{d+1}\pi^{d/2}\Gamma(d/2)$, $u_{\rho \rho}$, $v_z$
and $v_{\rho \rho'}$ stand, respectively, for $\partial_\rho^2 u$, $\partial_z v$ and $\partial_\rho \partial_{\rho'
}v$ evaluated for fields that minimize $u_k$ (the subscript $k$ has
been dropped for simplicity), and $m_{p,p'}^d(w,w')$ are dimensionless
threshold functions described in~\cite{berges02}. In
eq.(\ref{eq_etab}) we have set for clarity $T_k=T=0$, but the same FPs
are reached when $T>0$ provided (as indeed found) that
$\theta=\theta_{k\to0}>0$. As discussed previously~\cite{tarjus04},
breakdown of DR is associated with the appearance of a strong enough
nonanalycity in the renormalized 2-replica potential as the two
replica fields become equal. This nonanalycity is related to the
presence of metastable states in the renormalized random potential
generated along the RG flow.

We have numerically integrated the flow equations for $u_k$, $v_k$,
and $\eta_k$ for a variety of initial conditions and determined in
this way the FPs and the associated stability behavior (and critical
exponents). To cover the whole $(N, d)$ diagram for continuous values
of $N$ and $d$ with a tractable computational effort, we have used an
additional approximation that consists in expanding $u_k$ and $v_k$
around the field configuration $\rho_{\Min,k}$:
$u_k=u_{2}(\rho-\rho_{\Min,k})^2$,
$v_k=v_{00}(z)+v_{10}(z)(\rho+\rho'-2\rho_{\Min,k})+
v_{20}(z)(\rho+\rho'-2\rho_{\Min,k})^2+v_{02}(z)(\rho-\rho')^2$. One
should point out that the present flow equations reproduce all 1-loop
perturbative results in the appropriate region of the $(N, d)$ plane,
including the FRG equation at first order in
$\epsilon=d-4$~\cite{tarjus04}.

\begin{figure}[htbp]
  \includegraphics[width=0.53 \linewidth]{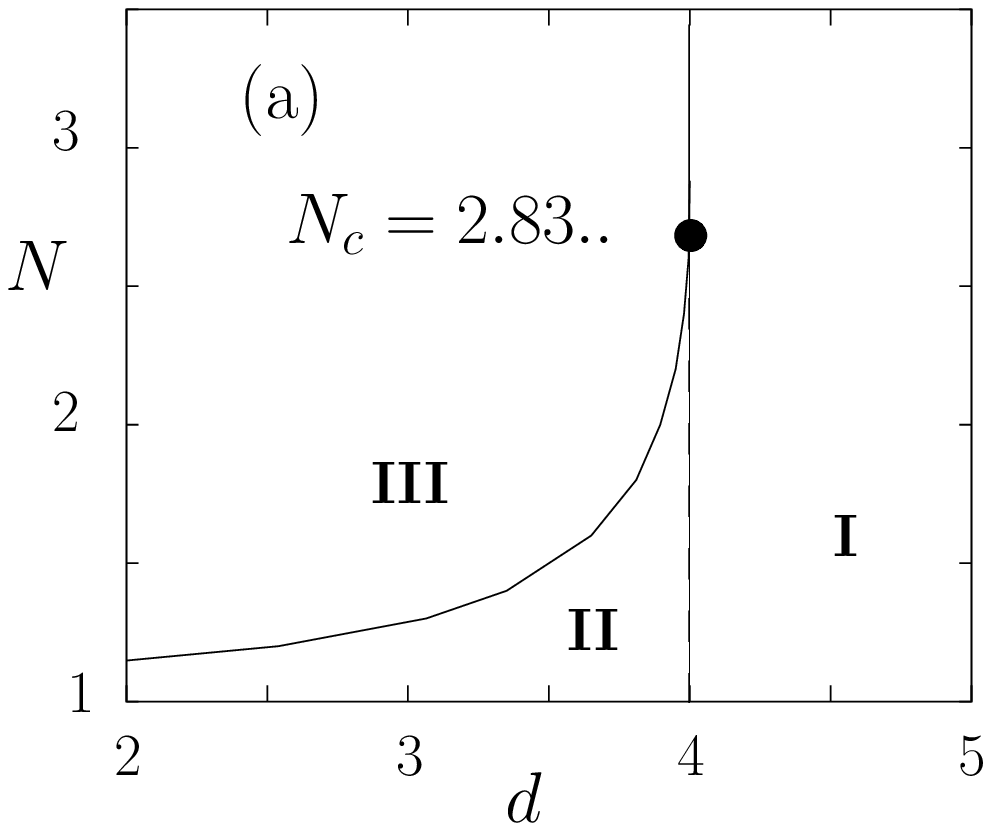}\hfill
  \includegraphics[width=0.47 \linewidth]{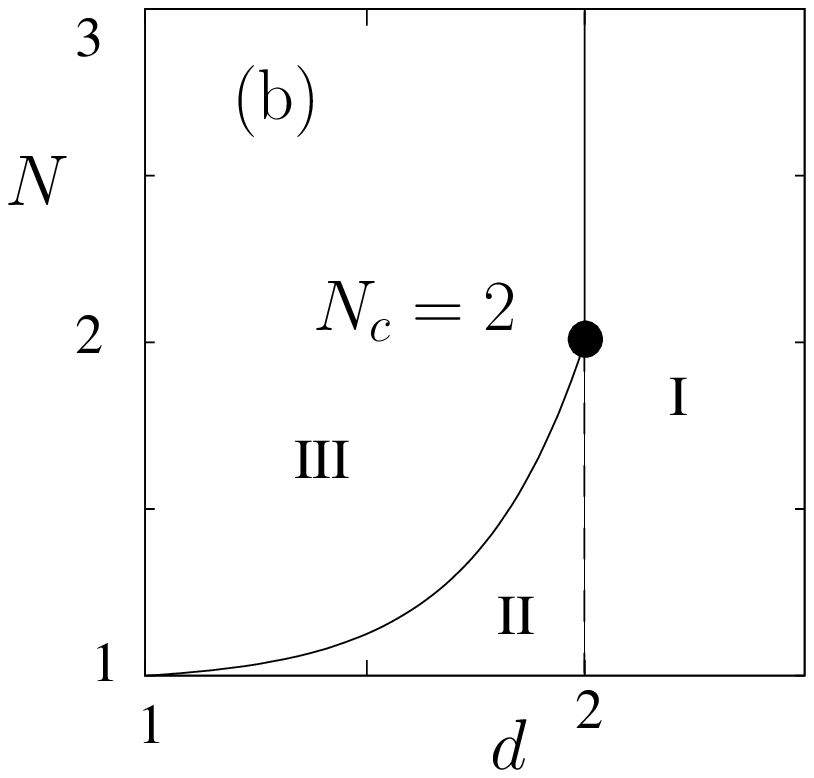}
   \caption{Predicted phase diagram for (a) the RF$O(N)$M
     (nonperturbative FRG) and (b) pure $O(N)$ model (sketch). Regions
     I and II correspond to transitions to a ferromagnetic and a QLRO
     phase, respectively.}
  \label{fig_nc}
\end{figure}
The central results of our study are summarized on the phase diagram
displayed in Fig.~\ref{fig_nc}a. In region III there are no phase
transitions, the RF$O(N)$M always stays disordered. In region I
there is a transition to a ferromagnetic phase at a critical point
governed by a zero-temperature nonanalytic FP. The
nonanalycity is strong enough (a "cusp") for breaking the DR
predictions; however, when approaching a critical line $N_{c,DR}(d)$
(not shown here, but starting from $N=18$ for $d=4+\epsilon$ and going
to $d \simeq5$ when $N=1$\cite{tarjus04}), the critical exponents
continuously tend to their DR value. Above the line, the stable (more
precisely, once unstable) FP is nonanalytic, but the
nonanalycity is now too weak to break DR. (FPs with a "cusp"
can still be found, but they are unstable in several directions and
correspond therefore to multicritical points, unreachable from generic
initial conditions.)

\begin{figure}[htbp]
  \centering
  \includegraphics[width=0.8\linewidth]{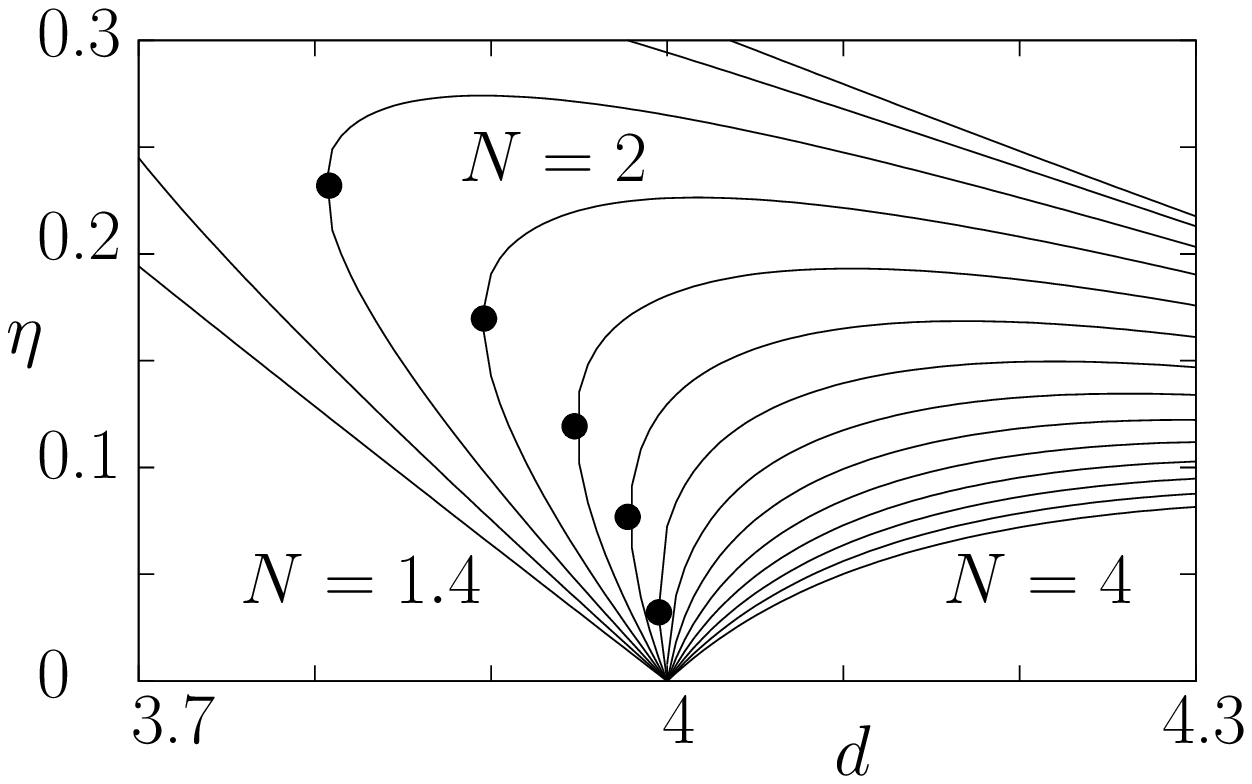}
  \caption{Prediction for the $d$-dependence of the anomalous
  dimension $\eta$ for different values of $N$ (by
  steps of $0.2$). Below $N_c=2.83...$ and for $d<4$ two nontrivial
  FPs are found and coalesce for $d=d_{lc}(N)$ (dots).}
  \label{fig_eta}
\end{figure}

\begin{figure}[htbp]
  \centering
  \includegraphics[width=0.8\linewidth]{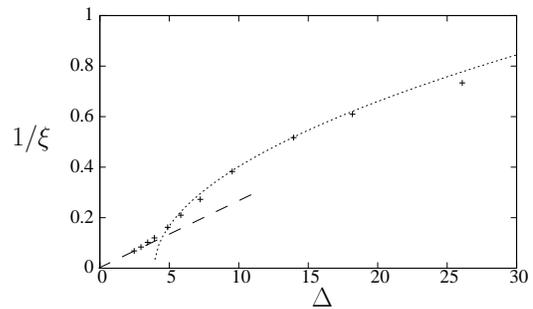}
  \caption{Inverse correlation length vs disorder strength for the
    3-$d$ RF$XY$M. The crossover around $\Delta=5$ is the remain
    of the FPs found above $d_{lc}$. $\Delta$ parametrizes the initial
    conditions of the RG flow and $1/\xi$ is estimated as in Ref.
    \cite{delamotte04}.}
  \label{fig_correlation}
\end{figure}

Finally, in region II we find two zero-temperature nonanalytic FPs:
one is attractive and describes a QLRO phase and the other one is once
unstable and governs the transition between the disordered and the
QLRO phases. The exponent $\eta$ characterizing the power-law decay of
the connected correlation function is shown in Fig.~\ref{fig_eta} for
a range of values of $N$.  One can see that for $N$ less than a
critical value $N_{c}=2.83...$ and for $d<4$, which corresponds to
region II, the two FPs coalesce for a value $d_{lc}(N)$ which then
determines the lower critical dimension below which no phase
transition is observed.  The most striking outcome is that, contrary
to what is usually believed, no QLRO, \textit{i.e.}, no Bragg glass
phase, exists in $d=3$ for the RF$XY$M: indeed,
$d_{lc}(N=2)\simeq3.8$. Our study of the RG flow for $d=3$ and $N=2$
shows signatures of a "ghost" nontrivial FP, presumably lying on the
imaginary axis not too far from the physical plane of coupling
constants (which could explain the behavior found in Monte Carlo
simulation\cite{gingras96}, see Fig. \ref{fig_correlation}), but the
flow goes at large distance to the trivial disordered FP. We shall
come back to this point later.  Note that an estimate of the
uncertainty of the present nonperturbative but approximate RG
treatment is given by considering the point in Fig.~\ref{fig_nc}a for
$d=2$: in this (probably most unfavorable) case, the theory predicts
$N_{lc}\simeq1.15$ instead of the exact result $N_{lc}=1$. As shown
for the pure $O(N)$ model\cite{ballhausen04}, this could be improved
by solving the full first order of the derivative expansion.

It is instructive to contrast the behavior of the RF$O(N)$M with that
of the pure model. The phase diagram of this latter, derived from the
solution of phenomenological RG equations~\cite{cardy80} and from
known exact results, is shown in Fig.~\ref{fig_nc}b. Regions I, II and
II are the exact counterparts of those in Fig.~\ref{fig_nc}a. Indeed,
and although never mentioned, QLRO exists in the pure model even aside
from the special KT transition for $N=2$ and $d=2$. However, it occurs
in an unphysical region, $1<N<2$, $1<d<2$ ! It remains that the
topology of the RF$O(N)$M phase diagram is very similar to that of the
pure model. Again, there is no exact DR-like property, the lower
critical dimension $d_{lc}(N)$ being shifted by an $N$-dependent value
between the two models.

Another intriguing question raised by this similarity is the possible
equivalence between the ($N_c=2.83...$, $d=4$) point of the RF$O(N)$M
and the ($N_c=2$, $d=2$) point of the pure model (compare Figs
\ref{fig_nc}a and b). Our nonperturbative FRG solution of the former
shows a behavior very reminiscent of the KT transition, but the
approximation used is unable to rigorously locate a line of FPs as
would occur for a KT transition. As studied in detail for the pure
$d=2$ XY model \cite{gersdorff01,berges02}, a line of ``almost'' FPs
is found: this is enough to observe a behavior almost
indistinguishable from the KT scenario, but not satisfying to
\textit{prove} the existence of a KT transition. To resolve this
matter, we have considered the perturbative FRG analysis around $d=4$.
The 1-loop $\beta$-function (or rather functional) for the
renormalized 2-replica potential (to which our nonperturbative
treatment exactly reduces near $d=4$) can be shown to vanish
identically in $d=4$ when $N=N_c=2.83...$ for arbitrary values of the
disorder strength, much like it does in the pure $d=2$ XY model for
arbitrary values of the temperature; but what about the 2-loop
$\beta$-function? To answer this question we have calculated
(independently from the authors of the recent ref.
[\onlinecite{ledoussal05}]) the $\beta$-functions of the RF$O(N)$M
around $d=4$ to 2 loops, starting with the nonlinear-sigma model
associated with Eq.~(\ref{eq_action_replicated}). The calculation
follows the treatment developed for the pure model~\cite{brezin76},
with the bare action expressed in terms of renormalized dimensionless
quantities,
\begin{equation*}
  \label{eq_action_sigma}
  \begin{split}
    S[\{\vect \pi_a\}]=-\frac{k^{d}}{2Z_T^2 t^2}\int_x \sum_{a,b=1}^n
   [ R(Z_\pi \vect \pi_a\cdot\vect
    \pi_b+\sigma_a \sigma_b)+\delta[R]] \\
    +\frac{k^{d-2}}{2Z_T t}\int_x
    \sum_{a=1}^n[Z_\pi(\partial\vect\pi_a)^2+ (\partial\sigma_a)^2-2 h
    \frac{Z_t}{\sqrt Z_\pi}\sigma_a]
  \end{split}
\end{equation*}
where the ($N-1$)-component fields $\vect\pi_a$ ($\vect
\pi_a=Z_\pi^{-1/2}\vect \Pi_a$) characterize the fluctuations in the
directions orthogonal to the external field $h$ ($h=Z_\pi^{1/2}
Z_T^{-1} H$) while $\sigma_a=\sqrt{1-Z_\pi \vect \pi_a^2}$; $t=k^{d-2}
Z_T^{-1} T$ and $R(z)$ is the renormalized 2-replica potential that we
seek to relate to the bare term $R_0(z)=\Delta z$ via the counterterms
$\delta[R]$, such that $R_0=k^{4-d}(R+\delta[R])$. We then carry out
the perturbation expansion around the gaussian theory characterized by
the propagator $G^{(0)\mu\nu}_{ab}(q)=t Z_Tk^{2-d}(Z_\pi q^2+
Z_\pi^{1/2}Z_T h/\sigma_a)^{-1}
(\delta_{\mu\nu}-\pi_a^\mu\pi_a^\nu)\delta_{ab} $ where
$\mu,\nu=1,\dots N-1$. The relevant interaction vertices are obtained
by taking up to six functional derivatives of the action with respect
to the $\vect \pi_a$ fields.  They involve derivatives of $R$, \ie{},
functions of $z$. To handle the resulting functional diagrams we have
followed the method developed for the FRG of disordered elastic
systems~\cite{ledoussal04}.  Calculations are quite intricate and will
be reproduced elsewhere.

The final expression for the $T=0$ $\beta$-function for $R(z)$ can be
put in the form $\partial_t R(z)=R'(1)\beta[r]=R'(1)( \epsilon
r(z)+R'(1)\beta_1[r]+R'(1)^2\beta_2[r])$, where $R'(1)$ is the
renormalized disorder strength and $r(z)=R(z)/R'(1)$; the 1-loop
$\beta_1[r]$ is given in refs~\cite{fisher85, tarjus04} and the 2-loop
contribution reads
\begin{equation*}
  \label{eq_2_boucles}
  \begin{split}
&2 \beta_2[r]= y(y r'''-3 z r''-r')^2(-y r''+z r'-1)-(N-2)\\&\Big[ y^2
r''^3-y(3 z r'+y-3)r''^2-2y(r'-z)r' r''+y r'^2+4r\Big]\\&
+a^2 \Big [(N+2) y
r''-(3N-2) z r'+8K(N-2)r\Big],
  \end{split}
\end{equation*}
where $y=1-z^{2}$ and $a=\lim_{(z\to 1^-)}(1-z^2)^{ 1/2}r''(z)$. One
can check that with the change of variable $z=\cos \phi$ and with
$K=2\gamma_a$, the above equation is identical to the result recently
obtained by Le~Doussal and Wiese~\cite{ledoussal05}.  As expected, DR
is recovered by setting the ``anomalous'' term $a$ to zero. The
coefficient $K$ in the above expression is left so far as an unknown:
it requires additional calculations (in progress) in order to fully
determine the 2-loop $\beta$-functions. However, its value is not
needed to answer the above question, \ie, to know if for
$N_c=2.83\dots$ and $d=4$ the $\beta$-function for $R(z)$ vanishes
identically to two loops for arbitrary disorder strength $R'(1)$. This
latter property is only true if both $\beta_1[r]$ and $\beta_2[r]$
vanish identically for the same function $r(z)$. One can check that
this is impossible, irrespective of the value of $K$. We can thus
conclude that the special point $(N_c=2.83...$, $d=4$) does {\em not}
correspond to a KT transition. Actually, as also obtained in our
nonperturbative approach (see Fig.\ref{fig_nc}a), the only FP found
for the $(N_c$, $d=4$) model is at zero renormalized disorder
strength, which allows a perturbative analysis of the two nonanalytic
fixed points for $d<4$ and $N<N_c$ \cite{ledoussal05}.

One further consequence of the absence of a KT transition is that,
contrary to what occurs in the pure model near ($N=2$, $d=2$), the
line of lower critical dimension approaches the $(N_c$, $d=4$) point
with an infinite slope (compare Figs~\ref{fig_nc}a and b). This
reinforces our finding that $d=3$ is safely below the lower critical
dimension of the RF$XY$M. Whether or not this conclusion implies that
no Bragg glass phase should exist in disordered high-$T_c$
superconductors is however an open question. The mapping from these
latter to the RF$XY$M is only valid in the low-T phase where an
essentially elastic description can be used. It may well be that the
mechanism by which QLRO is destroyed in the RF$XY$M, a mechanism that
does not explicitly invokes the presence of dislocations, is specific
to that model.


\begin{thebibliography}{33}
\expandafter\ifx\csname natexlab\endcsname\relax\def\natexlab#1{#1}\fi
\expandafter\ifx\csname bibnamefont\endcsname\relax
  \def\bibnamefont#1{#1}\fi
\expandafter\ifx\csname bibfnamefont\endcsname\relax
  \def\bibfnamefont#1{#1}\fi
\expandafter\ifx\csname citenamefont\endcsname\relax
  \def\citenamefont#1{#1}\fi
\expandafter\ifx\csname url\endcsname\relax
  \def\url#1{\texttt{#1}}\fi
\expandafter\ifx\csname urlprefix\endcsname\relax\def\urlprefix{URL }\fi
\providecommand{\bibinfo}[2]{#2}
\providecommand{\eprint}[2][]{\url{#2}}

\bibitem[{\citenamefont{Imry and Ma}(1975)}]{imry75}
\bibinfo{author}{\bibfnamefont{Y.}~\bibnamefont{Imry}} \bibnamefont{and}
  \bibinfo{author}{\bibfnamefont{S.~K.} \bibnamefont{Ma}},
  \bibinfo{journal}{Phys. Rev. Lett.} \textbf{\bibinfo{volume}{35}},
  \bibinfo{pages}{1399} (\bibinfo{year}{1975}).

\bibitem[{\citenamefont{Bricmont and Kupianen}(1987)}]{bricmont87}
\bibinfo{author}{\bibfnamefont{J.}~\bibnamefont{Bricmont}} \bibnamefont{and}
  \bibinfo{author}{\bibfnamefont{A.}~\bibnamefont{Kupianen}},
  \bibinfo{journal}{Phys. Rev. Lett.} \textbf{\bibinfo{volume}{59}},
  \bibinfo{pages}{1829} (\bibinfo{year}{1987}).
\bibinfo{author}{\bibfnamefont{J.~Z.} \bibnamefont{Imbrie}},
  \bibinfo{journal}{Phys. Rev. Lett.} \textbf{\bibinfo{volume}{53}},
  \bibinfo{pages}{1747} (\bibinfo{year}{1984}).

\bibitem[{\citenamefont{Larkin}(1970)}]{larkin70}
\bibinfo{author}{\bibfnamefont{A.~I.} \bibnamefont{Larkin}},
  \bibinfo{journal}{Sov. Phys. JETP} \textbf{\bibinfo{volume}{31}},
  \bibinfo{pages}{784} (\bibinfo{year}{1970}).

\bibitem[{\citenamefont{Aizenman and Wehr}(1989)}]{aizenman89}
\bibinfo{author}{\bibfnamefont{M.}~\bibnamefont{Aizenman}} \bibnamefont{and}
  \bibinfo{author}{\bibfnamefont{J.}~\bibnamefont{Wehr}},
  \bibinfo{journal}{Phys. Rev. Lett.} \textbf{\bibinfo{volume}{62}},
  \bibinfo{pages}{2503} (\bibinfo{year}{1989}).

\bibitem[{\citenamefont{Nattermann}(1998)}]{nattermann98}
\bibinfo{author}{\bibfnamefont{T.}~\bibnamefont{Nattermann}},
  \emph{\bibinfo{title}{Spin glasses and random fields}}
  (\bibinfo{publisher}{World scientific, Singapore}, \bibinfo{year}{1998}), p.
  \bibinfo{pages}{277}.

\bibitem[{\citenamefont{Feldman}(2001)}]{feldman01}
\bibinfo{author}{\bibfnamefont{D.~E.} \bibnamefont{Feldman}},
  \bibinfo{journal}{Int. J. Mod. Phys. B} \textbf{\bibinfo{volume}{15}},
  \bibinfo{pages}{2945} (\bibinfo{year}{2001}).

\bibitem[{\citenamefont{Giamarchi and {Le Doussal}}(1998)}]{giamarchi98}
\bibinfo{author}{\bibfnamefont{T.}~\bibnamefont{Giamarchi}} \bibnamefont{and}
  \bibinfo{author}{\bibfnamefont{P.}~\bibnamefont{{Le Doussal}}},
  \emph{\bibinfo{title}{Spin glasses and random fields}}
  (\bibinfo{publisher}{World scientific, Singapore}, \bibinfo{year}{1998}), p.
  \bibinfo{pages}{277}.
\bibinfo{author}{\bibfnamefont{T.}~\bibnamefont{Giamarchi}} \bibnamefont{and}
  \bibinfo{author}{\bibfnamefont{P.}~\bibnamefont{{Le Doussal}}},
  \bibinfo{journal}{Phys. Rev. Lett.} \textbf{\bibinfo{volume}{72}},
  \bibinfo{pages}{1530} (\bibinfo{year}{1994});
  \bibinfo{journal}{Phys. Rev. B} \textbf{\bibinfo{volume}{52}},
  \bibinfo{pages}{1242} (\bibinfo{year}{1995}).

\bibitem[{\citenamefont{Feldman}(2000{\natexlab{a}})}]{feldman00}
\bibinfo{author}{\bibfnamefont{D.~E.} \bibnamefont{Feldman}},
  \bibinfo{journal}{Phys. Rev. B} \textbf{\bibinfo{volume}{61}},
  \bibinfo{pages}{382} (\bibinfo{year}{2000}{\natexlab{a}});
  \bibinfo{journal}{Phys. Rev. B} \textbf{\bibinfo{volume}{62}},
  \bibinfo{pages}{5364} (\bibinfo{year}{2000}{\natexlab{b}}).

\bibitem[{\citenamefont{Blatter et~al.}(1994)\citenamefont{Blatter, Feigel'man,
  Geshkenbern, Larkin, and Vinokur}}]{blatter94}
\bibinfo{author}{\bibfnamefont{G.}~\bibnamefont{Blatter}},
  \bibinfo{author}{{\it et al.}}, 
\bibinfo{journal}{Rev. Mod. Phys.}
  \textbf{\bibinfo{volume}{66}}, \bibinfo{pages}{1125} (\bibinfo{year}{1994}).
\bibinfo{author}{\bibfnamefont{T.}~\bibnamefont{Nattermann}} \bibnamefont{and}
  \bibinfo{author}{\bibfnamefont{S.}~\bibnamefont{Scheidl}},
  \bibinfo{journal}{Adv. Phys.} \textbf{\bibinfo{volume}{49}},
  \bibinfo{pages}{607} (\bibinfo{year}{2000}).

\bibitem[{\citenamefont{Gingras and Huse}(1996)}]{gingras96}
\bibinfo{author}{\bibfnamefont{M.~J.~P.} \bibnamefont{Gingras}}
  \bibnamefont{and} \bibinfo{author}{\bibfnamefont{D.~A.} \bibnamefont{Huse}},
  \bibinfo{journal}{Phys. Rev. B} \textbf{\bibinfo{volume}{53}},
  \bibinfo{pages}{15193} (\bibinfo{year}{1996}).

\bibitem[{\citenamefont{Fisher}(1997)}]{fisher97}
\bibinfo{author}{\bibfnamefont{D.~S.} \bibnamefont{Fisher}},
  \bibinfo{journal}{Phys. Rev. Lett.} \textbf{\bibinfo{volume}{78}},
  \bibinfo{pages}{1964} (\bibinfo{year}{1997}).
\bibinfo{author}{\bibfnamefont{J.}~\bibnamefont{Kerfeld}},
  \bibinfo{author}{{\it et al.}},
  \bibinfo{journal}{Phys. Rev. B} \textbf{\bibinfo{volume}{55}},
  \bibinfo{pages}{626} (\bibinfo{year}{1997}).

\bibitem[{\citenamefont{Tarjus and Tissier}(2004)}]{tarjus04}
\bibinfo{author}{\bibfnamefont{G.}~\bibnamefont{Tarjus}} \bibnamefont{and}
  \bibinfo{author}{\bibfnamefont{M.}~\bibnamefont{Tissier}},
  \bibinfo{journal}{Phys. Rev. Lett} \textbf{\bibinfo{volume}{93}},
  \bibinfo{pages}{267008} (\bibinfo{year}{2004}).

\bibitem[{\citenamefont{Berges et~al.}(2002)\citenamefont{Berges, Tetradis, and
  Wetterich}}]{berges02}
\bibinfo{author}{\bibfnamefont{J.}~\bibnamefont{Berges}},
  \bibinfo{author}{{\it et al.}},
  \bibinfo{journal}{Phys. Rep.} \textbf{\bibinfo{volume}{363}},
  \bibinfo{pages}{223} (\bibinfo{year}{2002}).

\bibitem{delamotte04}
\bibinfo{author}{\bibfnamefont{B.}~\bibnamefont{Delamotte}},
  \bibinfo{author}{{\it et al.}}, 
  \bibinfo{journal}{Phys. Rev. B} \textbf{\bibinfo{volume}{69}},
  \bibinfo{pages}{134413} (\bibinfo{year}{2004}).

\bibitem[{\citenamefont{Ballhausen et~al.}(2004)\citenamefont{Ballhausen,
  Berges, and Wetterich}}]{ballhausen04}
\bibinfo{author}{\bibfnamefont{H.}~\bibnamefont{Ballhausen}},
  \bibinfo{author}{{\it et al.}}, 
  \bibinfo{journal}{Phys. Lett. B} \textbf{\bibinfo{volume}{582}},
  \bibinfo{pages}{144} (\bibinfo{year}{2004}).

\bibitem[{\citenamefont{Cardy and Hamber}(1980)}]{cardy80}
\bibinfo{author}{\bibfnamefont{J.~L.} \bibnamefont{Cardy}} \bibnamefont{and}
  \bibinfo{author}{\bibfnamefont{H.~W.} \bibnamefont{Hamber}},
  \bibinfo{journal}{Phys. Rev. B} \textbf{\bibinfo{volume}{45}},
  \bibinfo{pages}{1217} (\bibinfo{year}{1980}).

\bibitem[{\citenamefont{v.~Gersdorff and Wetterich}(2001)}]{gersdorff01}
\bibinfo{author}{\bibfnamefont{G.}~\bibnamefont{v.~Gersdorff}}
  \bibnamefont{and}
  \bibinfo{author}{\bibfnamefont{C.}~\bibnamefont{Wetterich}},
  \bibinfo{journal}{Phys. Rev. B} \textbf{\bibinfo{volume}{64}},
  \bibinfo{pages}{054513} (\bibinfo{year}{2001}).

\bibitem[{\citenamefont{{Le Doussal} and Wiese}()}]{ledoussal05}
\bibinfo{author}{\bibfnamefont{P.}~\bibnamefont{{Le Doussal}}}
  \bibnamefont{and} \bibinfo{author}{\bibfnamefont{K.~J.} \bibnamefont{Wiese}},
  \bibinfo{note}{cond-mat/0510344}.

\bibitem[{\citenamefont{Br\'ezin and Zinn-Justin}(1976)}]{brezin76}
\bibinfo{author}{\bibfnamefont{E.}~\bibnamefont{Br\'ezin}} \bibnamefont{and}
  \bibinfo{author}{\bibfnamefont{J.}~\bibnamefont{Zinn-Justin}},
  \bibinfo{journal}{Phys. Rev. B} \textbf{\bibinfo{volume}{14}},
  \bibinfo{pages}{3110} (\bibinfo{year}{1976}).

\bibitem[{\citenamefont{{Le Doussal} et~al.}(2004)\citenamefont{{Le Doussal},
  Wiese, and Chauve}}]{ledoussal04}
\bibinfo{author}{\bibfnamefont{P.}~\bibnamefont{{Le Doussal}}},
  \bibinfo{author}{{\it et al.}},
  \bibinfo{journal}{Phys. Rev. E} \textbf{\bibinfo{volume}{69}},
  \bibinfo{pages}{026112} (\bibinfo{year}{2004}).

\bibitem[{\citenamefont{Fisher}(1985)}]{fisher85}
\bibinfo{author}{\bibfnamefont{D.~S.} \bibnamefont{Fisher}},
  \bibinfo{journal}{Phys. Rev. B} \textbf{\bibinfo{volume}{31}},
  \bibinfo{pages}{7233} (\bibinfo{year}{1985}).
\bibinfo{author}{\bibfnamefont{D.~E.} \bibnamefont{Feldman}},
  \bibinfo{journal}{Phys. Rev. Lett.} \textbf{\bibinfo{volume}{88}},
  \bibinfo{pages}{177202} (\bibinfo{year}{2002}).

\end{thebibliography}
 \end{document}